\documentclass[a4paper]{jpconf}
\usepackage{graphicx}
\begin{document}
\title{Deformation effects in Giant Monopole Resonance}

\author{J Kvasil$^1$, V O Nesterenko$^2$,
A Repko$^{1}$, D Bozik$^{1}$, W Kleinig$^{2,3}$ and \\ P -G Reinhard$^{4}$}

\address{$^1$ Institute of Particle and Nuclear Physics MFF UK, Charles University, CZ-18000 Prague 8, Czech Republic}
\address{$^2$ Laboratory of Theoretical Physics, Joint Institute for Nuclear Research, Dubna, Moscow region, 141980, Russia}
\address{$^3$ Technical University of Dresden, Institute for Analysis, D-01062, Dresden, Germany}
\address{$^4$ Institute of Theoretical Physics II, University of Erlangen, D-91058, Erlangen, Germany}

\ead{kvasil@ipnp.troja.mff.cuni.cz}

\begin{abstract}
The isoscalar giant monopole resonance (GMR) in Samarium isotopes
(from spherical $^{144}$Sm to deformed $^{148-154}$Sm) is
investigated within the Skyrme random-phase-approximation  (RPA)
for a variety of Skyrme forces. The exact RPA and its separable
version (SRPA) are used for spherical and deformed nuclei,
respectively. The quadrupole deformation is shown to yield two
effects: the GMR broadens and attains a two-peak structure due to
the coupling with the quadrupole giant resonance.
\end{abstract}

\section{Introduction}

During last decades, the GMR remains to be a subject on intense
studies (see \cite{Colo08,Av13,Stone14} for recent reviews and
discussions) as it provides a valuable information on the nuclear
incompressibility \cite{Bl80}. Unlike early explorations, the
present theoretical analysis is mainly done within the
self-consistent mean field models (SC-MFM) \cite{Ben03,Vre05}, in
particular those based on Skyrme forces
\cite{Ben03,Skyrme,Vau,En75}. A variety of experimental data
becomes available, see e.g. \cite{It03,Li07,Pa12,Yo04,Uc04,Yo13}.

Despite an impressive theoretical and experimental effort, some
GMR problems are not yet resolved. For example, the GMR
experimental data in spherical nuclei, Sn/Cd isotopes \cite{Li07,
Pa12} from one side and Pb/Sm isotopes \cite{Yo04, Uc04} from
another side, cannot be simultaneously reproduced by any SC-MFM
\cite{Colo08,Av13,Ve12,Ca12,AN13}. It is not yet clear if this is
caused by a poor theoretical description (e.g., too rough
treatment of the pairing impact \cite{Kh09,Ve12, Ca12, AN13}) or
by a lack of accuracy of experimental data
\cite{It03,Li07,Pa12,Yo04,Uc04,Yo13}. Note that two main
experimental group measuring GMR in $(\alpha,\alpha')$ reaction,
Texas A\&M University (TAMU) \cite{Yo04,Yo13} and Research Center
for Nuclear Physics (RCNP) at Osaka University
\cite{It03,Li07,Uc04}, indeed sometimes provide noticeably
different results.

The situation with GMR in deformed nuclei is even more vague.
Though there is an evident progress in experiment studies, e.g.
for soft Mo \cite{Yo13} and deformed Sm \cite{It03,Yo04} isotopes,
the SC-MFM calculations are yet at very beginning, which is
explained by a need to deal with a huge configuration space. So
the theoretical results on GMR in deformed nuclei are now mainly
reduced to early inconsistent studies based on phenomenological
mean fields \cite{Ki75,Za78,Ab80}. More than three decades ago,
two deformation effects have been predicted \cite{Ki75,Za78,Ab80}
and observed \cite{Bu80,Ga84} in GMR: i) broadening the resonance
and ii) onset of two-peak structure due to a coupling with the
$\mu=0$ branch of the giant quadrupole resonance (GQR). Obviously
it is worth to revisit these results by using a modern theoretical
framework, e.g. the Skyrme SC-MFM. This is just the aim of the
present study.

We consider GMR in a chain of Sm isotopes, from spherical
$^{144}$Sm to deformed $^{154}$Sm, within the Skyrme random-phase
approximation (RPA) approach.  For spherical nuclei,  the exact
RPA  method \cite{Rep} is used. For deformed nuclei, the separable
RPA (SRPA) \cite{Ne02,Ne06} is implemented. This method exploits
the self-consistent factorization of the residual interaction,
which drastically decreases the computational effort wile keeping
high accuracy of the calculations. SRPA was proved  as very
reliable and effective theoretical tool in investigation of
various electric \cite{Kl08,Kv14,NeRi} and magnetic \cite{Ve09}
giant resonances in both axially deformed and spherical nuclei. In
particular, SRPA was used for exploration of the impact of the
time-odd current density on the properties of GMR in $^{208}$Pb
\cite{NeRi}. In spherical nuclei, SRPA and exact RPA results are
about identical. Both methods are fully self-consistent.

The calculations employ  a wide set of Skyrme forces with various
isoscalar effective masses $m_0/m$: SkT6 \cite{To84}, SVbas
\cite{Kl09}, SkM*\cite{Ba82}, SGII \cite{VG81}, SLy6 \cite{Ch97}
and SkI3 \cite{Re95}.

\section{Theoretical framework}

The calculations are performed within  exact RPA \cite{Rep} for
spherical $^{144}$Sm and SRPA \cite{Ne02,Ne06} for soft and
axially deformed $^{148,150,152,154}$Sm. For every nucleus, the
equilibrium deformation $\beta_2$ is determined. Then, for the
isoscalar monopole transition operator $\hat{M} = \sum_i^A (r^2
Y_{00})_i$, the strength function
\begin{equation}
S(E0;E) = \sum_{\nu} E_{\nu} \: |\:\langle \nu | \hat{M} | 0
\:\rangle |^2 \: \xi_{\Delta}(E-E_\nu ) \label{4}
\end{equation}
with the Lorentz weight $\xi_{\Delta}(E-E_{\nu}) = \frac{1}{2 \pi}
\frac{\Delta}{(E-E_{\nu})^2 - \Delta^2/4}$ is computed. Here,
$|0\rangle$ is the ground state wave function, $|\nu\rangle$ and
$E_{\nu}$ are RPA states and energies, $\Delta$ is the averaging
parameter. The Lorentz weight roughly simulates smoothing effects
beyond RPA (coupling to complex configurations and escape widths)
and allows a convenient comparison of the calculated and
experimental strengths. In the present study, the averaging
$\Delta$= 2 MeV is found optimal. Note that the same averaging was
used for the giant dipole resonance \cite{Kl08}.

In SRPA calculations for deformed nuclei, the input (doorway)
operators $r^2 Y_{00}, \; r^4 Y_{00}, \; j_0(qr)Y_{00}$ (with
$q$=0.4, 0.6) and  $r^2 Y_{20}$  are used. Following the standard
SRPA procedure \cite{Ne02,Ne06}, the first operator is the
transition one. This operator favors surface excitations. Then
some next operators (with higher power and Bessel-function radial
dependence) are added to take into account the nuclear interior
motion.  Finally the quadrupole operator is included to take into
account the coupling to quadrupole excitations (to be omitted in
spherical nuclei). At 5 input operators, the rank of the SRPA
matrix is $5\times4$, which is much less than a huge rank of
conventional RPA matrices. As seen from Fig. 1, the set with 4
operators provides an excellent agreement between SRPA and exact
RPA results in spherical $^{144}$Sm.

In SRPA calculations for $^{148-152}$Sm, the pairing $\delta$-force
\begin{equation}
 V_{pair}(\vec{r},
\vec{r}^{\:'}) = V_0 \: \left[ \:1 - \eta \:\left(
\frac{\rho(\vec{r})}{\rho_0}\right)^{\gamma}
\:\right]\:\delta(\vec{r} - \vec{r}^{\:'}) \label{5}
\end{equation}
is used at the BCS level \cite{Ben00}. For SVbas, both
volume ($\eta$ = 0) and density-dependent surface ($\eta$ =1) pairing options are used.
Other Skyrme forces exploit the standard volume pairing.

The calculations use a large configuration space with
particle-hole (two-quasiparticle) energies up to 70-75 MeV. For
the force SVbas, the monopole strength summed at the relevant
energy interval 9-45 MeV exhausts the energy weighted sum rule $
\rm{EWSR} =\frac{\hbar^2}{2 \pi m} \:A\: \langle r^2 \rangle_0 $
 by 100-105$\%$,
depending on the isotope. The similar results are for other
forces.  The spurious mode lies at 2-7 MeV, i.e. safely below the
GMR concentrated at 10-20 MeV.
\begin{figure}[h]
\begin{minipage}{12pc}
\includegraphics[width=12pc]{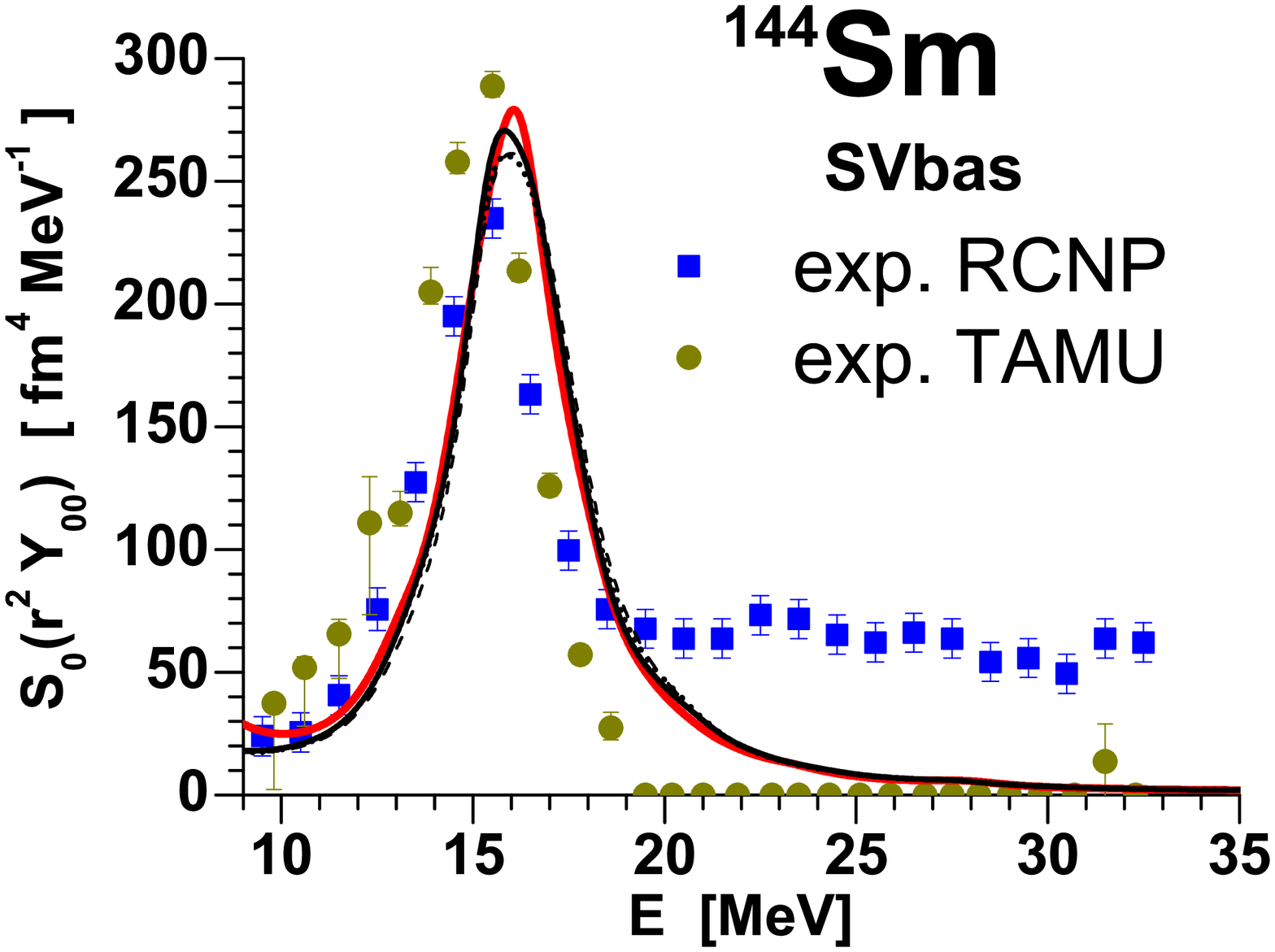}
\caption{\label{fig1} E0 strength function in the spherical
$^{144}$Sm calculated within full RPA (thick red line) and SRPA
(black lines). In SRPA, the input operators $f(r)Y_{00}$ with
radial dependence $r^2$ (dotted line), $r^2, r^4$ (dashed  line)
and $r^2, r^4,j_0(0.4r), j_0(0.6r)$ (thin line) are used. The
 RCNP \cite{It03} and TAMU \cite{Yo04} experimental data are depicted.}
\end{minipage}\hspace{2pc}%
\begin{minipage}{24pc}
\includegraphics[width=24pc]{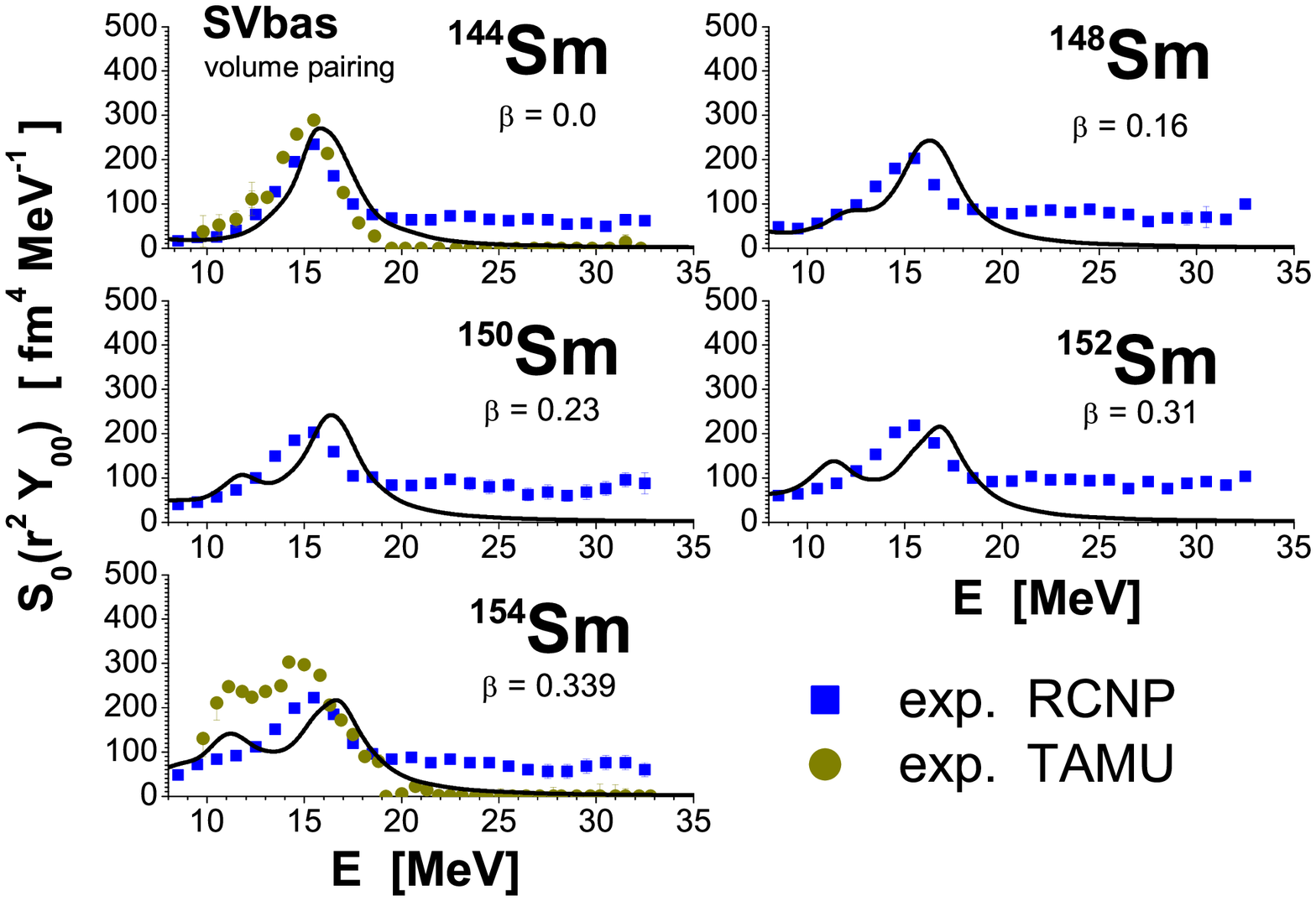}
\caption{\label{fig2} E0 strength functions in Sm isotopes,
calculated with SVbas force within full RPA ($^{144}$Sm) and SRPA
($^{148-154}$Sm), as compared  to RCNP \cite{It03} and TAMU \cite{Yo04}
experimental data. For every isotope, the deformation parameter
$\beta_2$ is shown.}
\end{minipage}
\end{figure}

\section{Results and discussion}

In Figure 1,  the E0 strengths calculated within full RPA and SRPA
( with different sets of input operators) are compared. It is seen
that RPA and SRPA strengths are about identical already for one
input operator $r^2 Y_{00}$, which indicates a high accuracy of
SRPA. This is partly caused by a simple one-peak form of the GMR.
Next three SRPA input operators are almost irrelevant here.
However, following our calculations,  they are necessary for
description of a more complicated GMR form in deformed nuclei.

Figure 1, as well as next figures, exhibit the RCNP \cite{It03}
and TAMU \cite{Yo04} experimental data. For convenience of the
comparison, the TAMU data, being initially presented in units of
the EWSR/E (where EWSR is the energy-weighted sum rule given in
Sec. 2 and E is the excitation energy), are transformed to units
$fm^4 MeV^{-1}$ used in RCNP. Following Fig. 1, the RCNP and TAMU
data give about the same GMR peak energy but considerably deviate
at energies above the GMR location, i.e. at $E>$ 19 MeV. Namely,
RCNP indicates a large and about uniform tail of E0 strength at 19
MeV$ < E <$ 33 MeV while TAMU gives here a vanishing strength (see
critical discussion of this discrepancy in  \cite{Yo04}). This
difference can affect determination of experimental GMR energy
centroids which are usually estimated through evaluation of sum
rules with different energy weights and thus depend on the chosen
energy interval.

As seen from Fig. 1, both full RPA and SRPA results well reproduce
the peak energy and width of the GMR (though the later is mainly
attributed to the proper choice of the average parameter
$\Delta$=2 MeV). For $E>$ 19 MeV, our calculations, in agreement
to TAMU data \cite{Yo04}, do not give any significant tail.

\begin{figure}[h]
\begin{center}
\includegraphics[width=21pc]{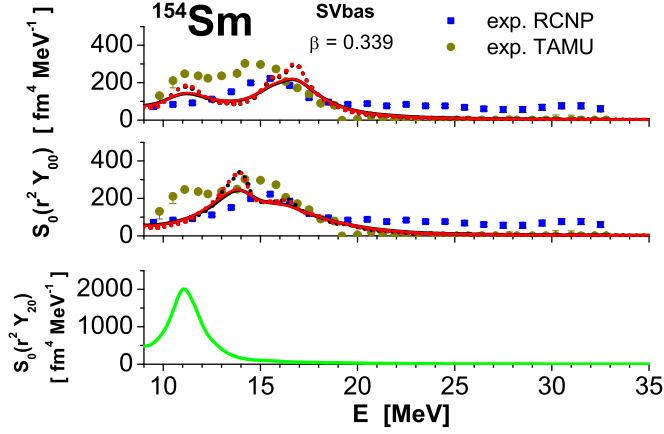}
\end{center}
\caption{\label{fig3} Isoscalar E0 (upper and middle panels) and
E2 (bottom panel) strength functions in deformed $^{154}$Sm
calculated within SRPA with the force SVbas. The E0 strength is
determined with (upper panel) and without (middle panel) coupling
to the quadrupole excitations. The E0 results obtained with the
volume (black curves) and surface (red curves) pairing as well as
with the averaging  $\Delta$=1 MeV (dotted curves) and 2 MeV
(solid curves), are compared. The RCNP \cite{It03} and TAMU
\cite{Yo04} experimental data are shown.}
\end{figure}
\begin{figure}
\begin{center}
\includegraphics[width=23pc]{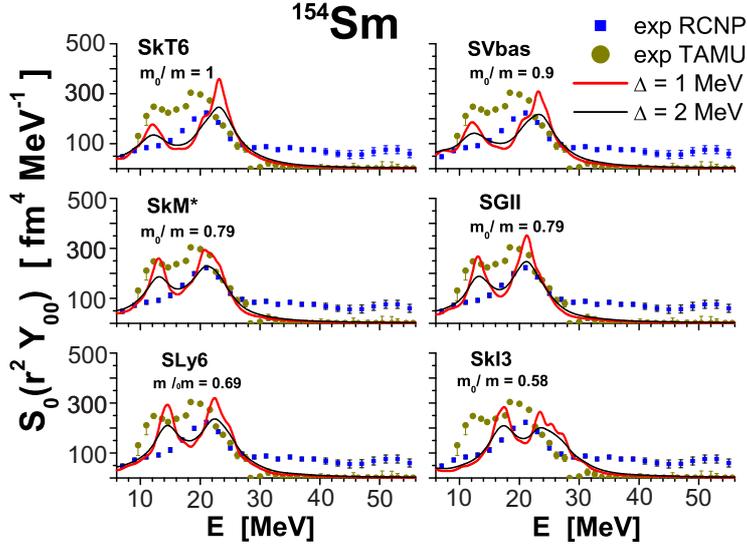}
\end{center}
\caption{\label{fig4} Isoscalar E0 strengths in $^{154}$Sm, calculated
for different Skyrme parametrizations (SkT6, SVbas, SkM*, SGII, SLy6 and SkI3)
and two values of the Lorents averaging parameter, $\Delta$ = 1 and 2 MeV, as
compared to  TAMU \cite{Yo04} and RCNP \cite{It03} experimental data. For each force,
the isoscalar effective mass $m_0/m$ is shown.}
\end{figure}

In Figure 2, the full RPA and SRPA results obtained with the force
SVbas are compared to TAMU \cite{Yo04} and RCNP \cite{It03} data
for the set of Sm isotopes. Parameters of the equilibrium
quadrupole deformation $\beta_2$ determined from the minimum of
the nuclear energy show that only $^{144}$Sm is spherical while
other isotopes are soft ($^{148}$Sm) or well deformed
($^{150-154}$Sm). The E0 strength is calculated within full RPA in
$^{144}$Sm and SRPA in $^{148-154}$Sm (with 5 input operators as
discussed above). It is seen that our calculations well reproduce
broadening of the GMR and onset of the GMR two-peak structure with
growth of the deformation from $^{144}$Sm to $^{154}$Sm. The
latter effect is caused by the coupling of E0 and E2 modes in
deformed nuclei \cite{Ki75,Za78,Ab80,Bu80,Ga84}. Note that the
two-peak structure in deformed Sm isotopes is observed in TAMU
\cite{Yo04} but not in RCNP \cite{It03} experiments, which ones
more signals on the essential difference between measurements of
these two groups.

The origin of the two-peak structure is demonstrated in Fig. 3 for
$^{154}$Sm. It is seen that the first peak takes place if E0-E2
coupling is included through the input operator $r^2Y_{20}$ (upper
plot) but is absent if the coupling is switched off (middle plot).
Moreover, the position of the first peak in E0 strength coincides
with position of $\mu=0$ branch of the quadrupole resonance,
exhibited at the bottom plot. The upper and middle plots also show
sensitivity of the results to the choice of pairing (volume vs
surface) and average parameter $\Delta$ (1 or 2 MeV). Though these
factors somewhat change the results, the qualitative picture
remains the same.

Finally in Fig.4, the GMR calculated in  $^{154}$Sm with different
Skyrme forces is shown. It is seen that the first GMR peak is
generally upshifted with decreasing the isoscalar effective mass
$m_0/m$. The better agrement is obtained for the forces with a
large $m_0/m$, from SkT6 to SGII. All the Skyrme forces give the
two-peak GMR structure. Our results for E0 strength at the GMR
region and above are in a general agrement with  TAMU \cite{Yo04}
data and do not correspond to RCNP \cite{It03} distributions. The
results with $\Delta$=1 and 2 MeV look qualitatively similar
though the larger averaging is more convenient for the comparison
to experiment. It is remarkable that changing $\Delta$ from 1 to 2
MeV practically does not affect the description of the GMR width
in deformed nuclei. The same was found for the giant dipole
resonance \cite{Kl08}.

\section{Conclusions}

The GMR in Sm isotopes, from spherical $^{144}$Sm to deformed
$^{148-154}$Sm,  was explored. We used exact RPA \cite{Rep} for
spherical $^{144}$Sm and separable RPA ( SRPA) \cite{Ne02,Ne06}
for deformed isotopes (both methods are fully self-consistent).
Various Skyrme forces (SkT6, SVbas, SkM, SGII, SLy6  and SkI3)
were involved. The SRPA calculations have shown distinct
deformation effects: i) broadening the GMR and ii) splitting of
the resonance into two peaks. The latter effect was shown to be
caused by the coupling between GMR and $\mu$=0 branch of the
quadrupole giant resonance. The obtained results are in a good
agreement with  TAMU experimental data \cite{Yo04} which support
the two-peak GMR structure. At the same time, they deviate from
RCNP \cite{It03} data which exhibit a one-bump GMR structure and
an impressive high-energy distribution of E0 strength. Certainly,
further exploration of GMR in Sm isotopes needs a harmonization of
available experimental data.

The calculations with different Skyrme forces give rather close
results though the forces with a large isoscalar effective mass
$m_0/m$ (from SkT6 to SGII) look more promising. In accordance to
previous studies, the volume and surface pairings give similar
effects in GMR.

\ack The work was partly supported by the DFG grant RE 322/14-1,
Heisenberg-Landau (Germany-BLTP JINR), and Votruba-Blokhincev
(Czech Republic-BLTP JINR) grants. P.-G.R. and W.K. are grateful
for the BMBF support under the contracts 05P12RFFTG and
05P12ODDUE, respectively. The support
of the Czech Science Foundation (P203-13-07117S) is appreciated.


\end{document}